\documentclass[journal]{IEEEtran}
\IEEEoverridecommandlockouts

\usepackage{cite}
\usepackage{amsmath,amssymb,amsfonts}
\usepackage{graphicx}
\usepackage[font=small,skip=1mm]{caption}
\usepackage{textcomp}
\usepackage{xcolor}
\usepackage{tabulary}
\usepackage{multirow}
\usepackage{mathtools}
\usepackage{booktabs}
\usepackage{siunitx}
\usepackage{etoolbox}
\AtBeginEnvironment{tabular}{\scriptsize} 
\begin{document}

\title{Temperature Dependent Reverse Recovery Characterization of SiC MOSFETs Body Diode for Switching Loss Estimation In a Half-Bridge}

\author{Debiprasad~Nayak,~\IEEEmembership{Student~Member,~IEEE,}
        Yakala~Ravi~Kumar,~\IEEEmembership{Student~Member,~IEEE,}
        Manish~Kumar,~\IEEEmembership{Student~Member,~IEEE,}
        Sumit~Pramanick,~\IEEEmembership{Member,~IEEE}}

\maketitle

\begin{abstract}
In a hard switched MOSFET based converter, turn-on energy losses is predominant in the total switching loss. At higher junction temperature the turn-on energy loss further increases due to the reverse recovery effect of the complementary MOSFETs body diode in a half-bridge configuration. Estimation of the switching loss under different operating conditions at an early design stage is essential for optimising the thermal design. Analytical switching loss models available in literature are generally used for estimating the switching losses, due to its accuracy and simplicity. In this paper, the inaccuracy in the reported loss models due to non inclusion of temperature dependent reverse recovery characteristics of body diode, is investigated. A structured method to determine the temperature-dependent switching loss of a SiC MOSFET in a half-bridge is presented. A simple methodology has been proposed to analyze the carrier lifetime's temperature dependencies of a SiC MOSFETs body diode. Device parameters from a $1.2kV/36A$ SiC MOSFETs datasheet are used for developing the loss model and experimental validation of the model. 
\end{abstract}


\section{Introduction}
SILICON carbide (SiC) MOSFETs have increasingly become popular as a replacement for Silicon (Si) based insulated gate bipolar transistors (IGBTs) due to its superior physical and electrical characteristics \cite{biela_sic_2011,millan_survey_2014}. Despite the improved performance, switching and conduction losses are the major loss contributor in a power semiconductor device. The efficiency of power conversion further reduces at high ambient temperature, as the junction temperature ($T_{j}$) increases with the increment of heatsink temperature ($T_{hs}$), which is common in automotive and grid-connected converters. The maximum ambient temperature in such applications typically ranges from $40^oC$ to $50^oC$ for grid-connected converters and $85^oC$ to $100^oC$ for automotive-grade converters. Therefore, an accurate estimation of semiconductor losses in such high temperature environment is essential for evaluating the efficiency and optimizing the cooling system for overall power density improvement. This work's primary focus is to model and accurately estimate the temperature-dependent switching losses of SiC MOSFETs in a half-bridge, which is the most commonly used building block for DC-DC/DC-AC converters.
 
\par
In the overall loss of a converter, the conduction loss can be calculated using the temperature-dependent parameter ($R_{DS}$), available in the datasheet. To accurately estimate the switching losses, different approaches are described in the literature\cite{mantooth_modeling_2015}. Among these, the most commonly used methods are physics-based models, numerical models, behavioural model and analytical models. In physics-based models, physical device data for SPICE modelling and field expertise are necessary \cite{kraus_physics-based_2016,potbhare_physical_2008}. Though physical models are accurate enough, it is hindered by the fact that many of its physical parameters are not present in the manufacturer datasheet. In numerical models, simulation tools like SILVACO and TCAD are being used. These simulation tools provide very accurate results, but these tools require material properties, device geometry and are very computationally intensive \cite{pushpakaran_electro-thermal_2013}. Behavioural models are the simplest model, and its solution is dependent on a couple of nonlinear equations which are based on curve fitting parameters \cite{alexakis_modeling_2013,merkert_characterization_2014}. In analytical models, device characteristics are divided into different segments, and governing equations are being derived  and parameterized based on datasheet parameters \cite{christen_analytical_2019,roy_analytical_2019,wang_characterization_2013}. In this work, the model presented is a combination of both behavioural and analytical model. 

\par
The main problem in estimating the switching losses comes from the modelling of the non-linear parasitic capacitances and the temperature-dependent reverse recovery loss of MOSFETs. Efforts have been made in \cite{yuancheng_ren_analytical_2006,wang_characterization_2013} to consider the non-linearity of the junction capacitance either by two-point or multi-point approximation through curve fitting approach. A voltage-dependent capacitor model is derived in \cite{Wang_Analytical_2019,roy_analytical_2019} and a charge equivalent capacitance model is given in \cite{christen_analytical_2019} to reduce the models' complexity. Though many methods are present in the literature, the effect of temperature on reverse recovery loss is not taken in to account for calculating the total loss of the MOSFETs. So at the higher ambient temperature, the above mentioned analytical techniques become erroneous. 

\par
In this paper, the analytical model proposed in \cite{christen_analytical_2019} has been improvised by including temperature dependencies. Estimation of turn-on loss has been emphasized more, as the increase in turn-off loss due to increment of temperature is negligible. The reverse recovery current due to the SiC MOSFETs body diode increases with the rise in $T_{j}$ due to the dependencies of charge carrier lifetime on temperature. Hence the temperature dependencies of the carrier lifetime and the drift region time constant has been considered and the parameter extraction process for these time constants has been described. Further the impact of parasitic capacitance on the reverse recovery process of the body diode has been discussed. During the switching transition,  $dV_{ds2}/dt$ and $dI_{d2}/dt$ of the device under test (DUT) are considered to be linear. 

\par 
This paper is structured as follows. Section II describes the model parameters used, and Section III describes the switching segments during turn-on with the effect of temperature on reverse recovery of the SiC MOSFETs body-diode. Section IV describes the experimental set-up and discusses experimental and analytical results. Section V concludes the paper.

\section{Model Parameters}
In the conventional MOSFETs loss estimation technique, a closed-form solution with a linear approximation of voltage and current is given for easy calculation \cite{erickson2007fundamentals,rohm,ti}. However, this method does not consider the effect of circuit parasitics and temperature, which results in inaccurate estimation of losses. Given the high switching rate of the SiC MOSFETs, consideration of the effect of parasitics and temperature becomes crucial to remove these inaccuracies. This section introduces to the circuit parasitics and the data extraction process from device datasheet for describing the governing equations during the switching intervals of the DUT. 

\subsection{Passive Parameters}
\begin{figure}[b!]
\setlength\baselineskip{-5mm}
\centering{\includegraphics[scale=0.38]{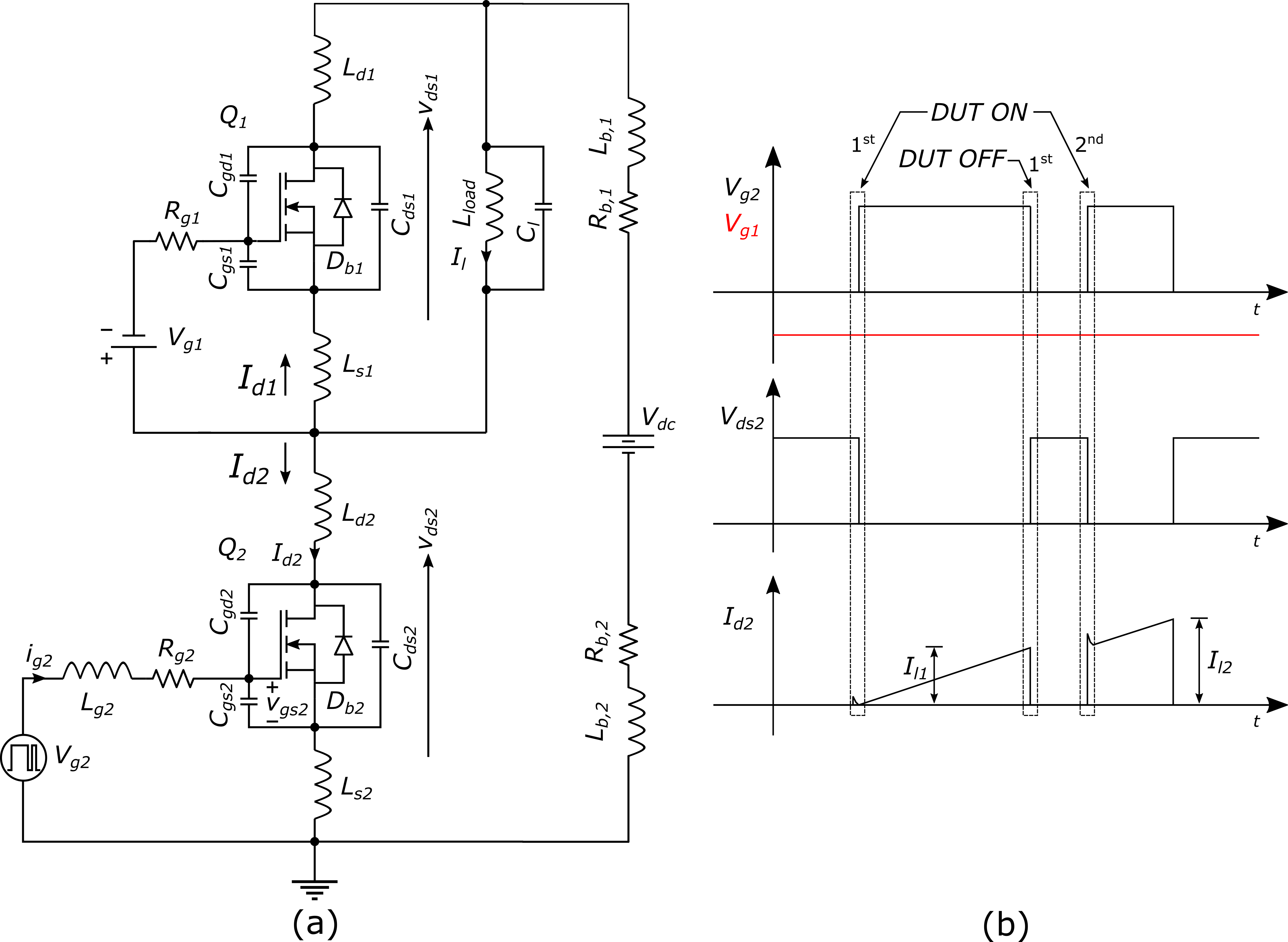}}
\caption{(a) Double pulse test circuit with circuit parasitics and (b) Switching waveforms in double pulse test}
\label{fig:dutwaveform}
\end{figure}

The circuit diagram, considering its parasitics, is shown in Fig. \ref{fig:dutwaveform}. The resistances included in this model are gate resistance $R_{g}$ and parasitic resistance present in the DC bus $R_{b}$. The $R_{b}$ considered here is the resistance from the nearest DC link capacitor to the switches which is the summation of positive rail DC bus resistance $R_{b,1}$ and negative rail DC bus resistance $R_{b,2}$. While switching, the damping for the $V_{ds}$ and $I_{d}$ oscillations depends on $R_{b}$. The parasitic inductances present in the circuit are $L_b$, $L_d$ and $L_s$. $L_b$ is the DC bus inductances which consists of $L_{b,1}$ and $L_{b,2}$. $L_d$, $L_s$ are the drain and source lead inductances of the SiC MOSFETs, respectively. The parasitic capacitances considered in this model are the gate-source capacitance $C_{gs}$, drain-source capacitance $C_{ds}$, gate-drain capacitance $C_{gd}$ of the SiC MOSFETs and the load capacitance $C_L$.

\begin{figure}[h!]
\centering{\includegraphics[scale=0.24]{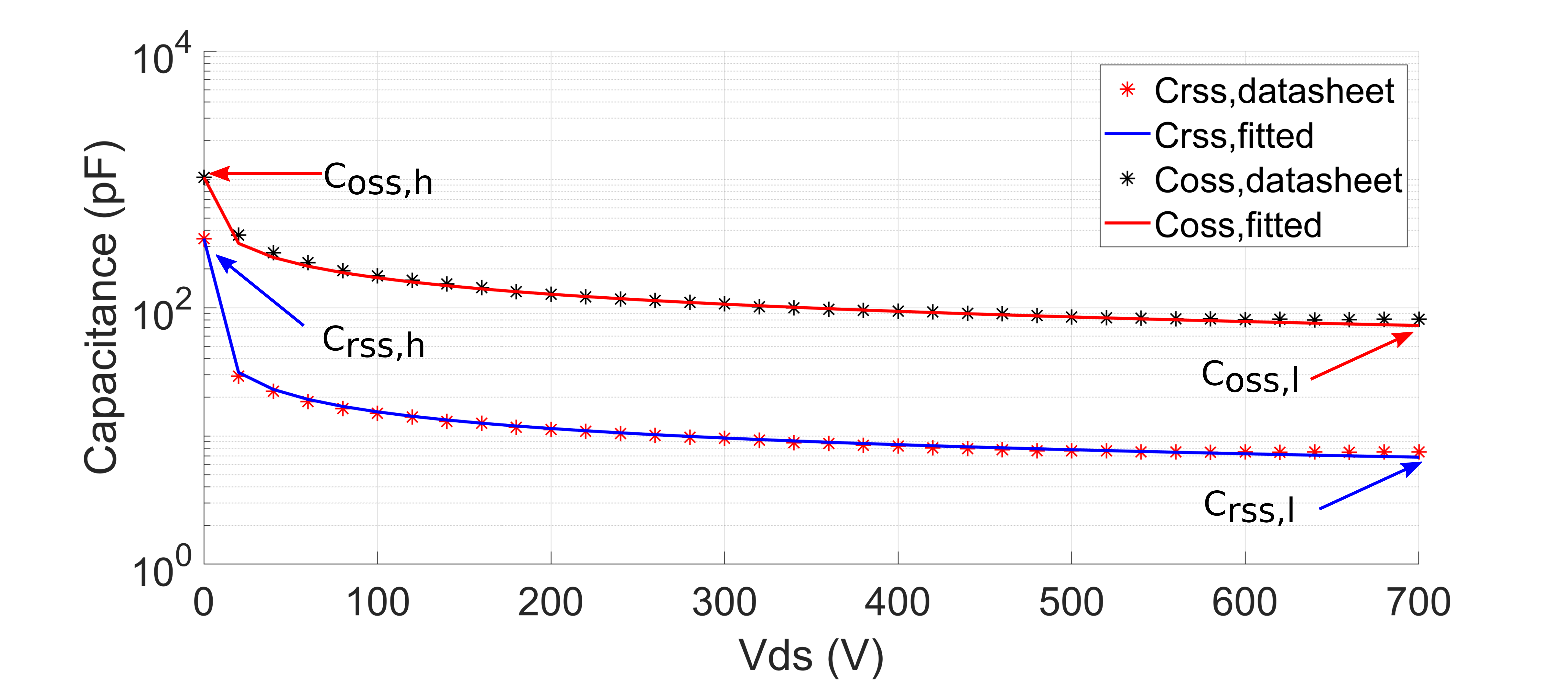}}
\setlength{\belowcaptionskip}{-5mm}
\caption{ Coss and Crss for CREE C2M0080120D SiC MOSFET and their fitted curves}
\label{fig:curvefit_cap}
\end{figure}

\par
Typically the capacitances values provided in the device datasheets are input, output and reverse transfer capacitances ($C_{iss}$, $C_{oss}$ and $C_{rss}$) which are voltage dependent and can be modelled as per eq. \ref{eqn:1} \cite{roy_analytical_2019}, and the average charge equivalent capacitances can be determined as per eq. \ref{eqn:2}

\begin{equation}
\label{eqn:1}
C_{x}(V_{\text{ds}})=\frac{C_{o,x}}{\sqrt{\displaystyle(1+\frac{V_{\text{ds}}}{a_{x}})}+b_{x}}
\end{equation}

\begin{equation}
\label{eqn:2}
C_{x,av}=\frac{1}{V_{DC}}\int\nolimits_{0}^{V_{DC}} C_{x}(v_{\text{ds}}) dv_{\text{ds}}
\end{equation}

\begin{center}
\hspace{-6.8em}
$\text{where}:x = iss, oss, rss$
\end{center}

\par 
$C_{o,x}$ is the capacitance value at $V_{ds}$ equals to zero. $a_{x}$ and $b_{x}$ can be found out by fitting the capacitance ($C_{x}$ vs $V_{ds}$) curves provided in datasheet \cite{noauthor_c2m0080120d_nodate}. $C_{iss}$ need not be fitted as its value remains almost constant with respect to $V_{ds}$ and can be directly taken from the datasheet. Fig. \ref{fig:curvefit_cap} shows the fitted curve of $C_{oss}$ and $C_{rss}$ of C2M0080120D SiC MOSFET \cite{noauthor_c2m0080120d_nodate}. From these capacitances the MOSFETs terminal capacitances can be calculated as:

\begin{equation}
\begin{aligned}
\label{eqn:3}
C_{gs} &= C_{iss} - C_{rss}\\
C_{ds} &= C_{oss} - C_{rss}\\
C_{gd} &= C_{rss}\\
\end{aligned}
\end{equation} 

\subsection{Transfer Equation}
During switching transition, the MOSFETs mainly operates in three regions: 1) cut-off region, 2) saturation region and 3) ohmic region. For estimating the switching energy loss of the MOSFET, channel current $i_{ch}$ during saturation region is considered. 

\begin{figure}[t]
\centering{\includegraphics[scale=1.2]{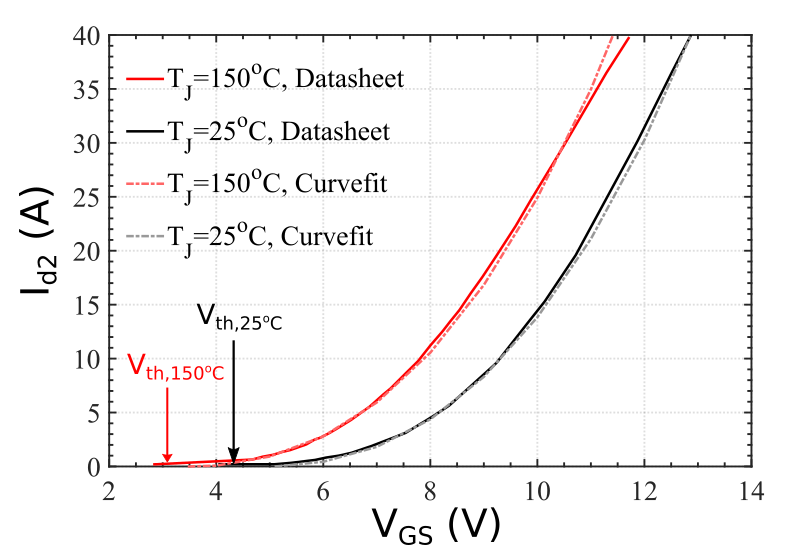}}
\setlength{\belowcaptionskip}{-5mm}
\caption{ Transconductance ($g_m$) characteristics of SiC MOSFET C2M0080120D}
\label{fig:gm}
\end{figure} 

In the saturation region the transfer characteristics of a MOSFET is given in its datasheet for a given temperature and can be expressed per eq. \ref{eqn:4}. The current transconductance factor $g_m$ depends on $i_{ch}$, and can be expressed as per eq. \ref{eqn:5}\cite{perret_power_2009}.  Due to the dependency on temperature, $g_{m}$ changes as $T_{j}$ increases Fig. \ref{fig:gm}. The effect of $T_{j}$ on $V_{th}$ can be expressed as per eq. \ref{eqn:6}. 

\begin{equation}
\label{eqn:4}
 i_{ch}=k_1{[v_{gs}-V_{th}(T_{j})]}^{x}+k_2
\end{equation}
\vspace{-0.5em} 
\begin{equation}
\label{eqn:5}
 g_{m}(i_{ch})=\sqrt[x]{\frac{k_1 i_{ch}^x}{i_{ch}-k_2}}
\end{equation}
\vspace{-0.5em} 
\begin{equation}
\label{eqn:6}
V_{th}(T_{j}) = aT_{j}^2 + bT_{j} + c 
\end{equation}

Here $a$, $b$, $c$, $k_1$, $k_2$ and $x$ are the curve-fitting parameters. The values of these parameters are $\num{29e-6}$, $\num{-15e-3}$, $4.9$, $\num{195e-3}$, $0$ and $2.5$ respectively for C2M0080120D.

\section{Switching Characterization and discussion}
This section introduces the switching process of the MOSFETs ($Q_1$ and $Q_2$) in a half-bridge configuration represented in Fig. \ref{fig:dutwaveform}(a). The lower MOSFET $Q_2$ is considered as the DUT.  

\par
The double-pulse test (DPT) has been divided into three parts as per Fig. \ref{fig:dutwaveform}(b). At the first turn-on event the voltage across $Q_1$, $Q_2$ are zero and $V_{dc}$ respectively. When the $Q_2$ is being turned on the parasitic capacitance $C_{oss,2}$ of $Q_2$ discharges and the $C_{oss,1}$ of $Q_1$ charges. The current that flows through the MOSFETs in the first turn-on event is due to the MOSFET's parasitic capacitances, which experiences a $dV_{ds}/dt$ across it. At the first turn-off event of $Q_2$, $I_{d2}$ starts to fall and $V_{ds2}$ starts to rise while charging its $C_{oss,2}$. The load current $I_{load}$ starts to transfer to the body-diode of $Q_1$ and $V_{ds1}$ starts to fall while discharging its $C_{oss,1}$. In the second turn-on event of $Q_2$, $I_{load}$ starts transferring from $Q_1$ to $Q_2$ and during this, in addition to the capacitive current, reverse recovery current of $Q_1$ increases the turn-on loss of $Q_2$. This turn-on loss increases significantly with the increase in $T_{j}$ due to reverse recovery current of the body diode of $Q_1$, which has been explained in Section III-B. 

\begin{figure}[t]
\centering{\includegraphics[scale=0.55]{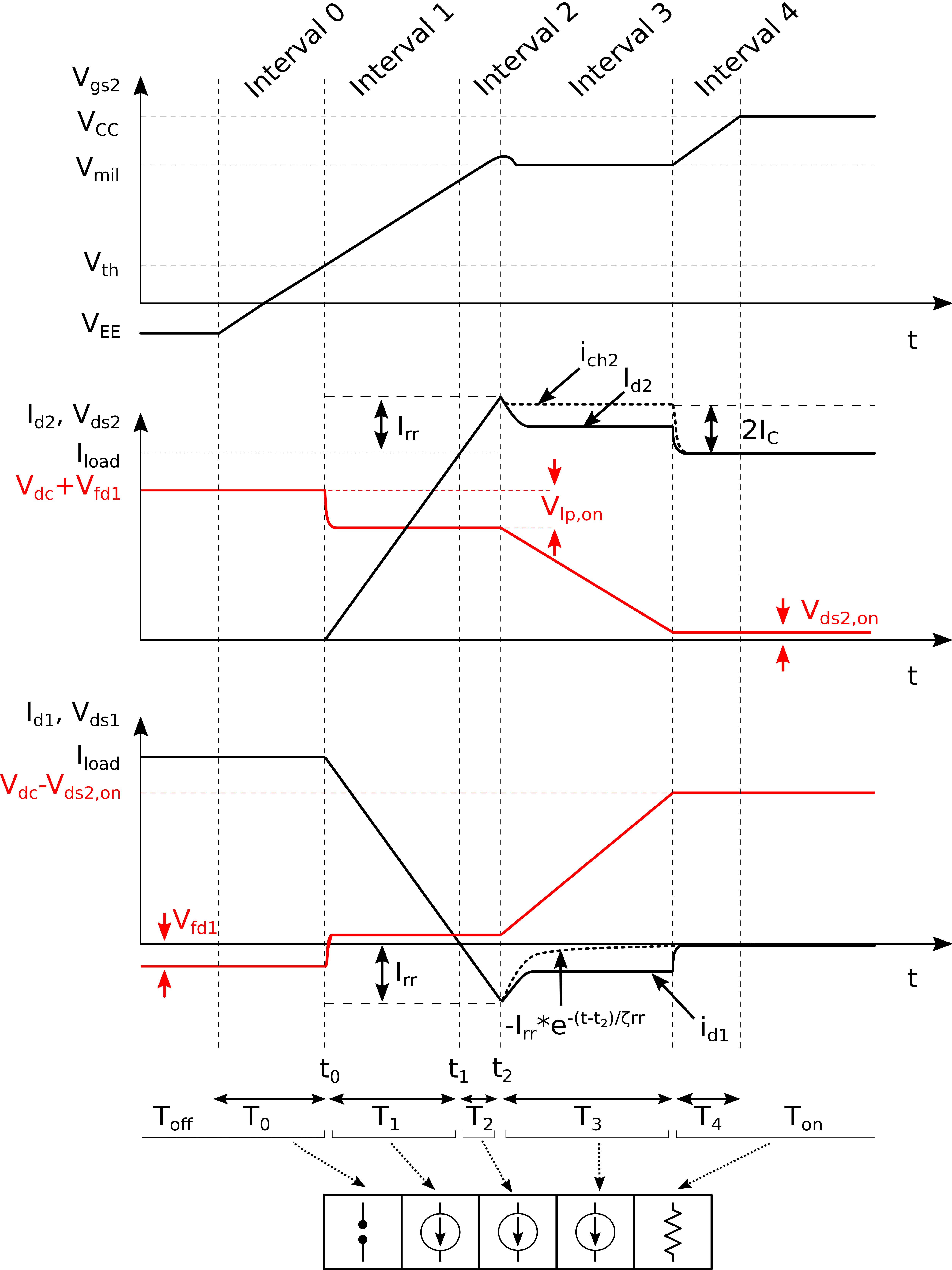}}
\setlength{\belowcaptionskip}{-5mm}
\caption{ Turn-on process MOSFET $Q_1$ and $Q_2$ in a half bridge }
\label{fig:turnon}
\end{figure}

\subsection{Turn-On Transient}
Under hard switching conditions, contribution of turn-on loss $E_{on}$ in the total MOSFET switching loss $E_{tot}$ is significant. However, $E_{on}$ increases as $T_{j}$ increases, so the effect of temperature must be included in the loss estimation. A detailed interval wise turn-on process can be seen from Fig. \ref{fig:turnon}.
  
\subsubsection{Interval 0 (Delay Time)}
\par
This stage starts with increase in gate voltage $V_{gs,2}$ from negative gate bias voltage $V_{EE}$ to threshold voltage $V_{th}$ with a time constant of $R_{g}(C_{gs}+C_{gd,h})$. This interval is known as turn-on delay time $t_{d,on}$. In this interval as $V_{gs,2}(t)<V_{th}$, there is no drain current $I_{d2}$ in MOSFET $Q_{2}$ and all the $I_{load}$ flows through the body diode of MOSFET $Q_{1}$. The voltage across the $V_{ds,2}$ remains at $V_{dc}+V_{fd1}$.

\vspace{-5mm}
\begin{equation}
\label{eqn:7}
\begin{split}
& v_{gs}(t) = V_{CC}+[V_{EE}-V_{CC}]e^{\dfrac{-t}{R_{g}(C_{gs}+C_{gd,h})}}\\
& T_{0} = R_{g}(C_{gs}+C_{gd,h})ln \dfrac{V_{CC}-V_{EE}}{V_{CC}-V_{th}(T_{j})}\\
& E_{on,0} = 0
\end{split}
\end{equation}

\subsubsection{Interval 1 (Current Rise Time)}
\par
In this interval as $v_{gs}(t)$ crosses the $V_{th}$, $Q_2$ channel current $I_{ch2}$ starts to rise to $I_{load}$. In this interval $I_{d2}$ remains equal to $I_{ch2}$ and the $V_{ds2}$ drops to $V_{dc}+V_{fd1}-V_{Lp,on}$. Voltage $V_{Lp,on}$ is the addition of voltage drops across $L_b$, $L_{d2}$ and $L_{s2}$. The dynamics of the gate voltage, time duration and the energy loss in this interval are expressed in eq. \ref{eqn:9}. The $dV_{gd}/dt$ and $dV_{ds}/dt$ does not change significantly in this interval, so for simplicity, these can be assumed to be zero.

\vspace{-1em}
\begin{equation}
\label{eqn:9}
\begin{split}
& v_{gs}(t) = V_{CC}+[V_{th}(T_{j})-V_{CC}]e^{\dfrac{-t}{R_{g}C_{gs}+L_{s}g_{m}(T_{j})}}\\
& T_{1} = |-(R_{g}C_{gs}+L_{s}g_{m}(T_{j}))ln \left( 1- \dfrac{I_{load}}{g_{m}(T_{j})(V_{CC}-V_{th}(T_{j}))} \right)|\\
& V_{Lp,on} = L_{p}\dfrac{I_{load}}{t_{1,2}}\\
& E_{on,1} = \dfrac{1}{2}T_{1}(V_{dc}+V_{fd1}-V_{Lp,on})I_{load}
\end{split}
\end{equation}

\subsubsection{Interval 2 (Storage Time)}
At this interval $I_{d2}$ continues to rise above $I_{load}$ to $I_{load}+I_{rr}$. Due to the reverse recovery of MOSFET $Q_1$, $V_{ds2}$ stays at $V_{dc}+V_{fd1}-V_{Lp,on}$. The circuit condition does not change from interval 1, but the time duration $T_{2}$ changes with respect to the $I_{load}$ and $T_{j}$. The detailed working process of MOSFETs body diode in this interval with $I_{load}$ and $T_{j}$ dependencies is explained in Section III-B. The energy loss of $Q_2$ in this interval can be expressed as eq. \ref{eqn:10}.

\vspace{-1em}
\begin{equation}
\label{eqn:10}
E_{on,2} =  (V_{dc}+V_{fd1}-V_{Lp,on})(I_{load} + \dfrac{1}{2}I_{rr})T_{2}
\end{equation}

\subsubsection{Interval 3 (Voltage Fall Time)}
After the current $I_{d1}(t)$ in the body diode of MOSFET $Q_1$ reaches to $I_{rr}$, it starts to block the voltage $V_{ds1}$ across it and simultaneously the $V_{ds2}$ across $Q_2$ starts to fall. Here as $V_{ds1}$ starts to rise, the parasitic capacitance $C_{oss1}$ starts to charge and the parasitic capacitance $C_{oss2}$ starts to discharge. The charging current of $C_{oss1}$ is termed as $I_{C1}$ and the discharging current of $C_{oss2}$ is termed as $I_{C2}$. Due to $C_{gd1}$ discharging, the voltage $V_{gs2}$ in this interval stays at miller plateau voltage $V_{mil}$. The net current which flows through the channel of MOSFET $Q_2$ is $I_{ch2}=I_{load}+I_{C1}+I_{C2}+I_{rr}e^{-(t-t_{2})/\tau_{rr}}$ (details about $\tau_{rr}$ is given in Section III-B). If both the MOSFET $Q_1$ and $Q_2$ are same then the parasitic capacitance across it can be considered to be same, so $I_{C1}$=$I_{C2}$=$I_{C}$. For calculating the energy loss in this interval, $i_{ch}$ has been divided in to two parts: first is due to $I_{load}+2I_{C}$ and  the second is due to $I_{rr}e^{-(t-t_{2})/\tau_{rr}}$. The capacitive current $I_{C}$ can be found from the quadratic equation given in \cite{christen_analytical_2019} as:

\begin{equation}
\label{eqn:11}
I_{C}= \mid\dfrac{-B_{1}+\sqrt{B_{1}^2 - 4A_{1}C_{1}}}{2A_{1}}\mid
\end{equation}
where:
\begin{align*}
&A_{1}= \dfrac{-2L_{s}}{Q_{C}R_{g}}&\\
&B_{1}= \dfrac{C_{gd,av}}{C_{gd,av}+C_{ds,av}} + \dfrac{2}{g_{m}(T_{j})R_{g}}&\\
&C_{1}= \dfrac{V_{CC}-V_{th}(T_{j})}{R_{g}}-\dfrac{I_{load}}{R_{g}g_{m}(T_{j})}&
\end{align*}

$I_{C}$ is being determined through numerical iterative method, as $g_{m}$ changes according to $i_{ch}$ and $T_{j}$. So the iteration process will continue until the error in the calculation of $I_{C}$ becomes negligible. $Q_{C}$ is the net charge stored in the parasitic output capacitances of $Q_1$ or $Q_2$ (as the identical MOSFETs will have same charge storing capacity) and can be written as eq. \ref{eqn:12}. The duration in which $V_{ds2}$ fall depends on the effective discharging time of the parasitic capacitance and is given as eq. \ref{eqn:13}.

\begin{equation}
\label{eqn:12}
Q_{C}= V_{dc}C_{oss2,av}
\end{equation}

\vspace{-0.5em}

\begin{equation}
\label{eqn:13}
T_{3}=\frac{Q_{C}}{I_{C}}
\end{equation}
 
The total energy loss in the MOSFET $Q_1$ in this interval can be written as

\begin{equation}
\label{eqn:14}
\begin{split}
E_{on,3}&=\dfrac{1}{2}(I_{load}+2I_{C})(V_{dc}+V_{fd1}-V_{Lp,on}-V_{ds2,on})T_{3}\\
 		 &+ (I_{load}+2I_{C})V_{ds2,on}T_{3}\\
 		 &+ \tau_{rr}I_{rr}[(V_{dc}+V_{fd1}-V_{Lp,on})(1-e^{(-T_{3}/\tau_{rr})})]\\
 		 &+ \tau_{rr}I_{rr}(V_{ds2,on}-V_{dc}-V_{fd1}+V_{Lp,on})\\
 		 &[(\tau_{rr}/T_{3})(1-e^{(-T_{3}/\tau_{rr})})-e^{(-T_{3}/\tau_{rr})}] 
		 \end{split} 
\end{equation}

\subsubsection{Interval 4}
This interval starts when MOSFET $Q_2$ enters from saturation region to ohmic region. The gate voltage $V_{gs2}$ starts rising from $V_{mil}$ to $V_{CC}$ and the $V_{ds2}$ remains at $V_{ds2,on}$, hence this interval can be neglected from the switching interval. The energy loss in this interval  can approximated as $E_{on,4}\approx V_{ds2,on}I_{d2}T_{4}$, where $T_{4} \approx 2R_{g}(C_{gs}+C_{ds2,l})$. 
From the five turn-on switching intervals, interval $0$, $4$ remains in the cutoff and ohmic regions. In interval $1$, $2$ and $3$, the SiC MOSFET remains in saturation as in this interval $V_{gs}> v_{th}$ and $V_{ds}> v_{gs}-V_{th}$.
Therefore  for calculating the total turn-on energy loss, switching intervals $1$, $2$ and $3$ are being considered and can be written as:
  
\begin{equation}
\label{eqn:int1.8}
E_{on,tot} = E_{on,1}+E_{on,2}+E_{on,3}
\end{equation}  

\subsection{Reverse Recovery Loss Model}
\begin{figure}[b]
\hspace{-5mm}
\includegraphics[scale=0.75]{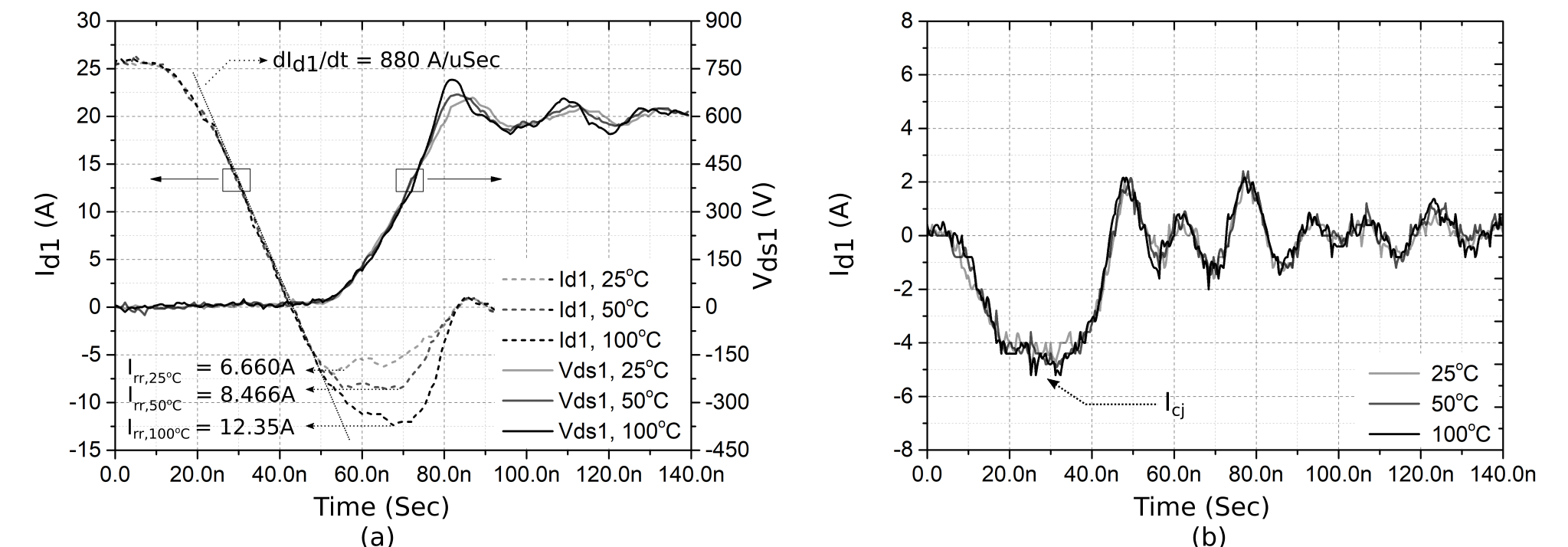}
\caption{ (a)Reverse recovery of C2M0080120D at different $T_{j}$ and (b)The capacitive displacement current $I_{cj}(t)$ of $Q_1$ under different $T_{j}$ at $1^{st}$ DUT on instance}
\label{fig:reverserecovery}
\end{figure}

This section describes the reverse recovery process of the body diode of the SiC MOSFET. The time period during this mode increases with respect to $T_{j}$ and $I_{load}$. So to estimate this duration, the power diode model from \cite{lauritzen_simple_1991,ma1997modeling} has been improved for better accuracy under different $T_{j}$ and further the impact of displacement current from parasitic capacitance has been discussed. For developing a comprehensive MOSFETs body diode model, the behaviour of the body-diode during turn-off needs to be analysed first. Fig. \ref{fig:reverserecovery}(a) shows the behaviour of a $36A/1.2kV$ SiC MOSFETs body diode during turn-off at different $T_{j}$.

\par
\begin{figure}[t!]
\centering{\includegraphics[scale=1]{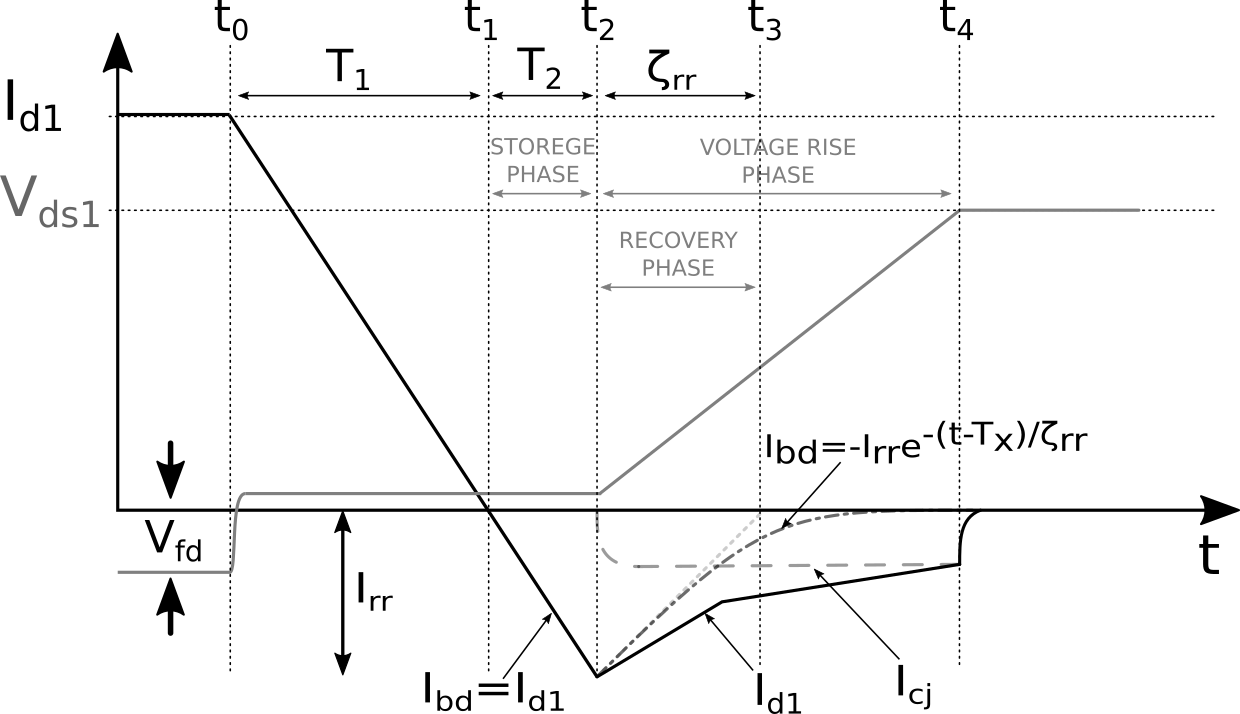}}
\setlength{\belowcaptionskip}{-5mm}
\caption{ Diode reverse recovery process}
\label{fig:reverse}
\end{figure}

The reverse recovery process starts when the current through the diode starts to flow in the negative direction. The rate at which body-diode current $I_{d1}(t)$ falls below zero, $dI_{d1}/dt$ depends on the circuit stray inductance $L_{p}$ and the applied reverse bias voltage $V_{ds1}$ across the MOSFET. As the diode current enters into the negative region, the excess charge carriers present in the junction starts to reduce while the $I_{d1}$ starts to fall towards the negative peak of the reverse recovery current $I_{rr}$. This region from $t_{1}$ to $t_{2}$ termed as storage phase. After the excess charge carriers become zero at $t_{2}$, depletion region starts to form and the reverse voltage across the diode starts to rise at the rate of $dV_{ds1}/dt$. Time period from $t_{2}$ to $t_{3}$ is termed as recovery phase (Fig. \ref{fig:reverse}).

\par
 After $t_2$, as $V_{ds1}$ starts rising towards $V_{dc}-V_{ds2,on}$, parasitic capacitance across the body-diode $C_{j} = C_{oss1}+C_{l}$ begins to charge and capacitive displacement current $I_{cj}(t)$ starts to flow. The magnitude of $I_{cj}(t)$ depends on the rate of $dV_{ds1}/dt$. Time duration from $t_{2}$ to $t_{4}$ is termed as the voltage-rise phase. The reverse-recovery characteristics in the voltage-rise phase depends upon two currents: actual body-diode reverse recovery current due to the depletion region formation $I_{bd}(t)$ and $I_{cj}(t)$. After $t_{4}$, $I_{d1}(t)$ reaches to its leakage current $I_{lk}$ level. The $I_{bd}(t)$ during the entire reverse-recovery phase can be written as per eq. \ref{eqn:ibd}. 
 
\begin{equation}
\label{eqn:ibd}
  I_{bd}(t)=\begin{cases}
    		I_{load}-\frac{V_{ds1}}{L_s}t , & \text{$t < t_2$}\\
    		-I_{rr}e^{-\frac{(t-t_2)}{\tau_{rr}}}-I_{cj}(t), & \text{$t_{2} \leq t \leq t_{4}$}\\
    		I_{lk}\approx 0, & \text{$t \geq t_{4}$}
  			\end{cases}
\end{equation}

\par
To further describe the dynamics of the body diode and to find out $I_{rr}$ for any operating conditions, three time constants are introduced as  \cite{lauritzen_simple_1991}: drift region transit time $T_{m}$, charge carrier life time $\tau_{c}$  and time constant of decay fall $\tau_{rr}$. These are inter related as per eq. \ref{eqn:tau}. From these time constants, $I_{rr}$ can be calculated by solving eq. \ref{eqn:irr_1}  numerically for $T_{2}$.

\begin{equation}
\label{eqn:tau}
\frac{1}{\tau_{rr}} = \frac{1}{\tau_{c}} + \frac{1}{T_{m}}
\end{equation}

\begin{equation}
\label{eqn:irr_1}
\begin{split}
T_{m} \left[ I_{load} -\dfrac{I_{load}}{T_1}(T_{1}+T_{2}) \right] = \\
\dfrac{I_{load}}{T_{1}} \tau_{c} \left[ -T_{2}+\tau_{c}-\tau_{c} e^{\dfrac{-(T_{1}+T_{2})}{\tau_{c}}}  \right]
\end{split}
\end{equation}

As the $T_{j}$ increases, these time constants tends to change. Hence the proposed extraction method in \cite{lauritzen_simple_1991} to find out $I_{rr}$ is not applicable for varying $T_{j}$. So an equivalent reverse recovery charge model is presented to find out the time period $T_2$. The total reverse recovery charge $Q_{rr}$ is calculated as per eq. \ref{eqn:qrr1}. This $Q_{rr}$ combines both the reverse recovery charge $Q_{rr}^{*}$, which is due to $I_{bd}(t)$ and the capacitive charge $Q_{cj}$ due to $I_{cj}(t)$ through parasitic capacitance $C_{j}$. As $T_{j}$ increases, the magnitude of $I_{rr}$ increases. Due to $I_{rr}$, $Q_{rr}^{*}$ also increases, but the $Q_{cj}$ remains constant, as $I_{cj}(t)$ does not change with respect to temperature Fig. \ref{fig:reverserecovery}(b). 

\begin{equation}
\label{eqn:qrr1}
 Q_{rr}=\int _{I_{d1}(t)<0} I_{bd}(t) dt 
\end{equation}

\begin{equation}
\label{eqn:qrr2}
 Q_{rr} = Q_{rr}^{*} + Q_{cj}
\end{equation}

\begin{equation}
\label{eqn:irr}
 I_{d1}(t) = I_{bd}(t) + I_{cj}(t)
\end{equation}

$I_{cj}(t)$ can be approximated as per eq. \ref{eqn:capcur} \cite{nayak_analysis_2020}, which remains constant for period $T_3$ time period. The capacitive charge $Q_{cj}$ can be found out from the first turn-on instant of DUT as per Fig. \ref{fig:reverserecovery}(b) by finding the area under the curve or from the datasheet parameters as per eq. \ref{eqn:capcur}. At this instant the net $Q_{rr}$ is only due to $Q_{cj}$, as the $I_{d1}(t)$ prior to this instant is zero. From Fig. \ref{fig:reverserecovery}(b) $Q_{cj}$ can be experimentally calculated as per eq. \ref{eqn:qrr1}.  

\begin{equation}
\label{eqn:capcur}
\begin{split}
I_{cj}(t) &=  C_{j}\frac{dV_{ds1}}{dt}\\
Q_{cj} &= \int\nolimits_{t_2}^{t_4} I_{cj}(t) dt 
\end{split}
\end{equation}

To find out the temperature dependencies of $Q_{rr}$, DPT at different $T_{j}$ needs to performed, from which $Q_{rr}^{*}$ can be calculated as per eq. \ref{eqn:qrr2}. In some manufacturer datasheet $Q_{rr}$ at two temperature are given which can be directly used to find out $Q_{rr}^{*}$. The time constant of decay fall $\tau_{rr}$ only depends on the $I_{rr}$, and the relationship between $\tau_{rr}$ and $Q_{rr}^{*}$ can be written as eq. \ref{eqn:trrqrr*}. Here it is assumed that after $\tau_{rr}$ time, the contribution of $I_{bd}(t)$ in $I_{d1}(t)$ is negligible as per Fig. \ref{fig:reverse}.

\begin{equation}
\label{eqn:trrqrr*}
Q_{rr}^{*} = 0.5*I_{rr}\left(\frac{1}{dI_{d1}/dt}I_{rr} + \tau_{rr}\right)
\end{equation}

To determine the  $\tau_{c}$, $\tau_{m}$ and $\tau_{rr}$ for any arbitrary operating conditions eq. \ref{eqn:tau}, eq. \ref{eqn:tc} and eq. \ref{eqn:tm} need to be solved numerically.

\begin{equation}
\label{eqn:tc}
I_{rr} = \frac{1}{2 \cdot dI_{d1}/dt}(\tau_{c}-\tau_{rr})\left(1-e^{{-T_1}/{\tau_c}}\right)
\end{equation}

\begin{equation}
\label{eqn:tm}
\frac{I_{rr}}{dI_{d1}/dt}(\tau_{c}+T_{m}) = \tau_{rr}^{2}\left(1-e^{{-T_1}/{\tau_c}}\right)
\end{equation}

\section{Results and discussion}
\subsection{Experimental Setup}
To study the effect of reverse recovery and switching energy loss at high $T_{j}$, the DPT has been carried out at $600V$ DC bus voltage and load current till $25A$. The experimental results are taken for the validation of the temperature dependent loss model proposed in this paper. A SiC MOSFET C2M0080120D rated at $1200V$ and $36A$ \cite{noauthor_c2m0080120d_nodate} is used for the half-bridge in the DPT setup. 

\par
For the gate-driver circuit, an isolated DC-DC converter with an output of $+20/-5V$ and an optoisolator $ACPL-337J$ are used. A total of $9.98 \Omega$ resistance is used as gate resistance $R_{g}$ for the MOSFETs. Low ESL DC link capacitors are being used near the half-bridge to reduce the DC bus inductance. A single layer air core inductor with an inductance of $475\mu H$ is used as a load inductor.

\par
\begin{figure}[ht!]
\centering{\includegraphics[scale=0.3]{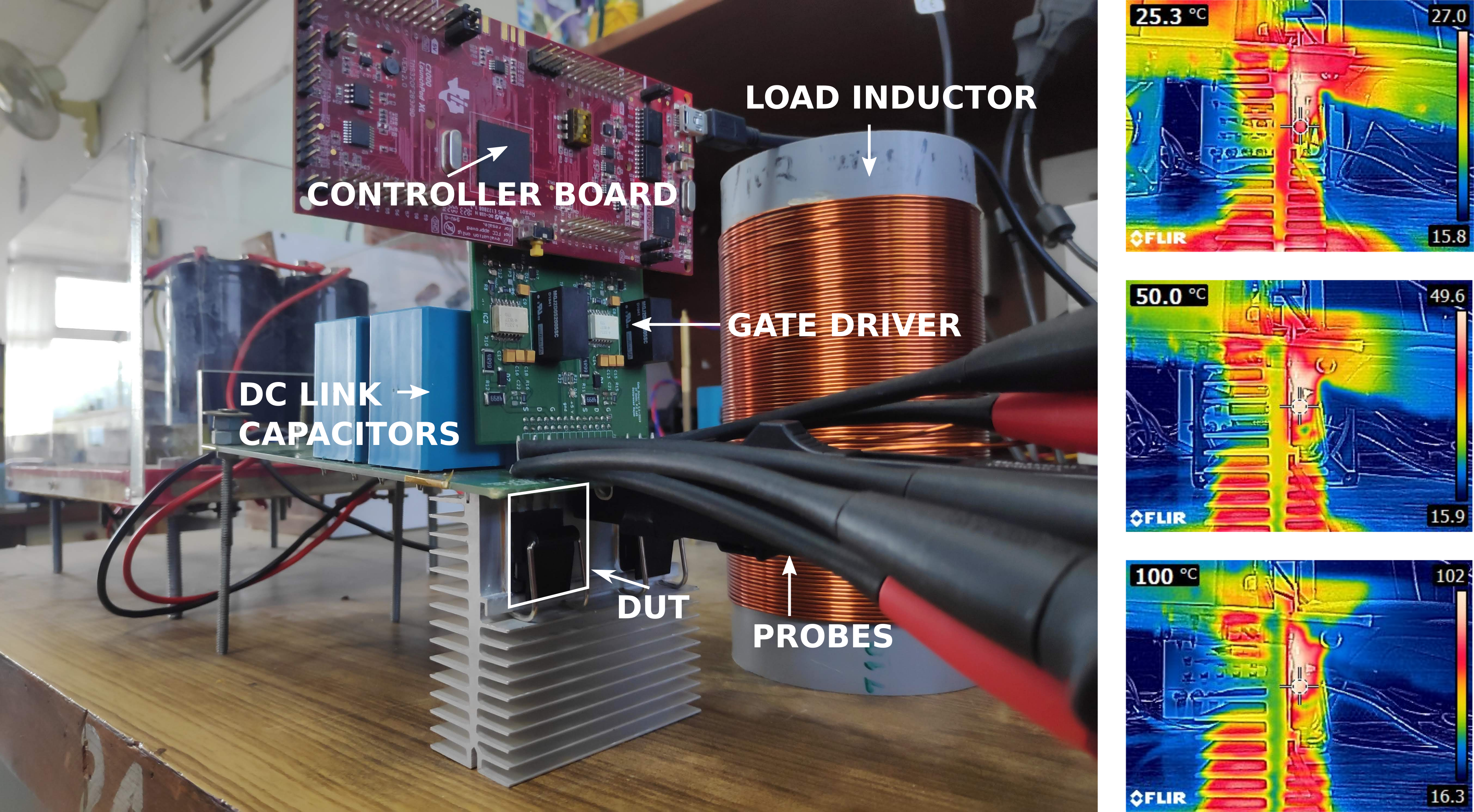}}
\setlength{\belowcaptionskip}{-3mm}
\caption{ Experimental test set-up with thermal images ($T_{hs}=25^oC, 50^oC \ \textrm{and} \ 100^oC$) of the DUT}
\label{fig:setup}
\end{figure}

A passive differential voltage probe (MDO3104) and a current probe (TCP0030A) from Tektronix are being used to measure the switching voltage and current transients. The heatsink was preheated before conducting the DPT at different temperatures until it reached a thermal equilibrium state to measure the switching performance. It is assumed that, under the thermal equilibrium state, the $T_{j}$ is equal to the $T_{hs}$. $T_{hs}$ is continuously monitored by a thermocouple temperature probe Fluke 80BK-A, and to further increase the accuracy of the reading, infrared thermal camera Flir E63900 is being used Fig. \ref{fig:setup}. 

\par 
Switching energy loss estimation is very sensitive to voltage-current timing misalignment. The propagation delay between the two probes is known as skew. So for accurate measurement, voltage and current probes need to be properly de-skewed. A resistive DPT has been conducted with a low inductive $30\Omega$ resistor to verify the propagation delay time for each measuring probes.

\subsection{Parameter Extraction}
\par
The parameters used in the loss model are extracted from the SiC MOSFETs datasheet and from the experimental results. 

\par
For calculating the transconductance $g_{m}(i_{ch})$, curve fit values based on the transfer characteristics from the datasheet of C2M00801120D are being used as per eq. \ref{eqn:5}. The datasheet and the fitted curves at two different $T_{j}$ are shown in Fig. \ref{fig:gm}. It can be seen that as the $T_{j}$ increases from $25^{o}C$ to $150^{o}C$, $V_{th}$ decreases from $4.5V$ to $3.3V$.

\par 
For extracting the non-linear capacitance values, eq. \ref{eqn:1} is used to find out the average capacitance values for saturation region. The maximum and minimum values of the capacitances are used for ohmic, cutoff regions, respectively. The curve fitting parameters and the extracted values are tabulated in Table \ref{tab:table1}\ \textrm{and} \ \ref{tab:table2}.

\par
\begin{figure*}[t]
\centering{\includegraphics[width=1.05\textwidth]{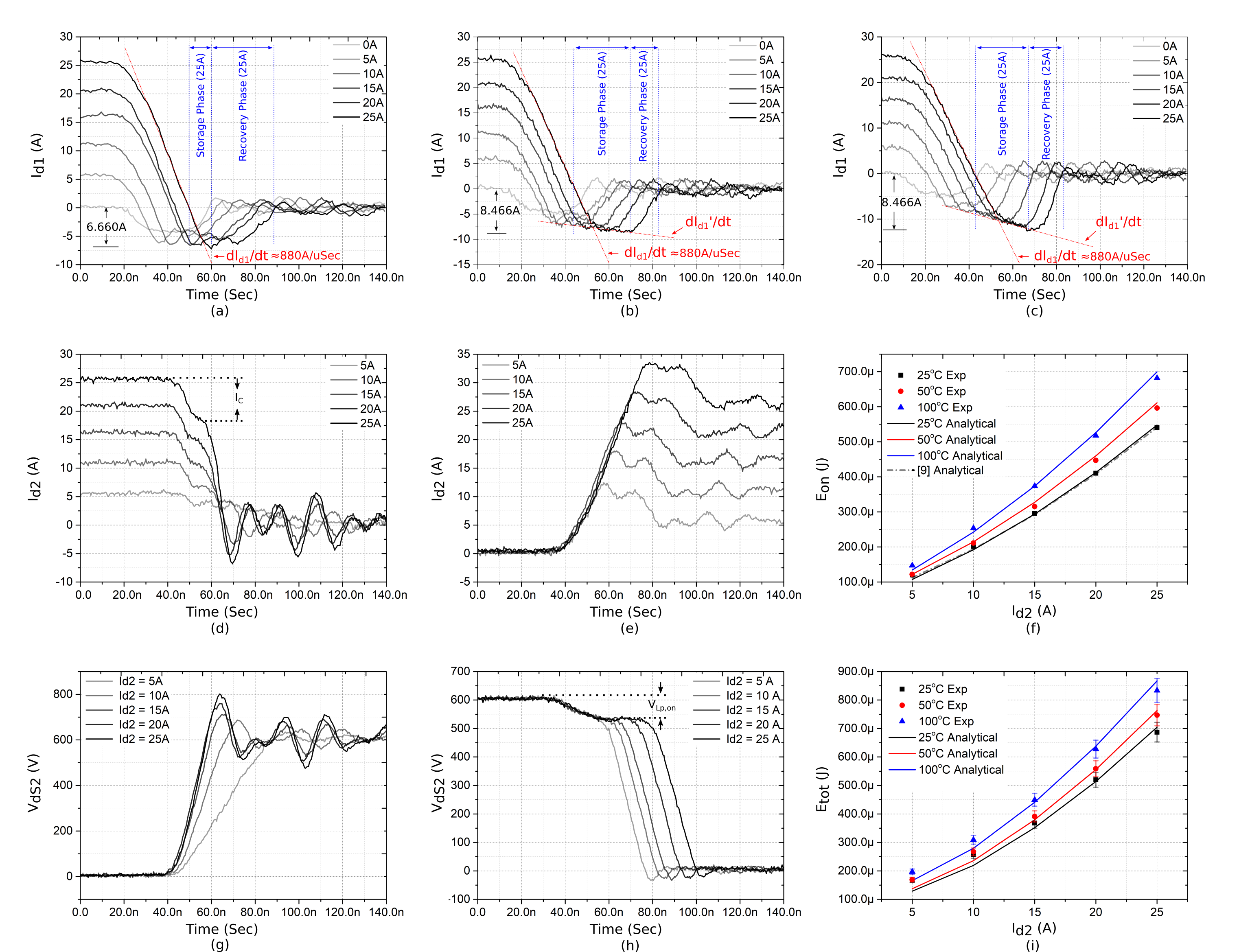}}
\setlength{\abovecaptionskip}{-2mm}
\setlength{\belowcaptionskip}{-3mm}
\caption{(a)Reverse recovery current $I_{d1}$ at $25^oC$, (b)Reverse recovery current at $I_{d1}(t)$ at $50^oC$, (c)Reverse recovery current at $I_{d1}(t)$ at $100^oC$, (d)$I_{d2}$ fall transition at $T_{j}=T_{hs}=25^oC$, (e)$I_{d2}$ rise transition at $T_{j}=T_{hs}=25^oC$, (f)Turn on energy loss at $T_{j}=T_{hs}=25^oC, 50^oC \ \textrm{and} \ 100^oC$, (g) $V_{ds2}$ rise transition at different load current under $T_{j}=T_{hs}=25^oC$, (h) $V_{ds2}$ fall transition at different load current under $T_{j}=T_{hs}=25^oC$, (i)Total energy loss at $T_{j}=T_{hs}=25^oC, 50^oC \ \textrm{and} \ 100^oC$ }
\label{fig:all}
\end{figure*}

 
The DC bus inductance is calculated from the voltage drop $V_{Lp,on}$ in $V_{ds2}$ due to the parasitic inductances from the turn-on switching waveform as shown in Fig. \ref{fig:all}(h). The total path inductance is found out to be $66.21 nH$ as per eq. \ref{eqn:9}. The influence of $L_{s2}$ is more on the switching transients and hence needs to be subtracted from the total path inductance and considered separately. In this work the value of $L_{s2}$ is considered to be $9nH$ \cite{zhang_integrated_2015,spice_model_c2m0080120d}. 

\par 
The extraction of fitting parameters for $Q_{rr}$ and $Q_{rr}^{*}$ are done at $T_{j}$ = $25^{o}C$, $50^{o}C$ and $100^{o}C$. The operating conditions are $V_{dc}$ = $600V$ and $I_{d1}$ = $25A$. Further to see the effect of load current on $dI_{d1}/dt$ during the fall time, $I_{d1}$ is varied from $0A$ to $25A$ as per Fig. \ref{fig:all}(a),(b) and (c). From all the test conditions, it is observed that $dI_{d1}/dt$ approximately remains constant at $880 A/\mu s$. It can be observed from Fig. \ref{fig:all}(a), at lower temperature ($T_{j}=25^oC$), during the storage phase the change in $dI_{d1}/dt$ is not significant with respect to load current. However, at higher temperature ($T_{j}=50^oC$ and $100^oC$) it can be observed from Fig. \ref{fig:all}(b) and (c), that $dI_{d1}/dt$ does not remain constant throughout the storage phase with respect to load current and tends to change before it reaches to peak reverse recovery current ($I_{rr}$). The change in the $dI_{d1}/dt$ during storage phase is termed as $dI_{d1}^1/dt$, and it increases with respect to $T_{j}$. Further as per Fig. \ref{fig:reverserecovery}(a), at higher $T_{j}$, $V_{ds1}$ starts to rise before the body-diode current reaches to $I_{rr}$. This phenomena increases the reverse recovery loss at a higher temperature. For simplicity in this work, it is assumed that $V_{ds1}$ rises after the $I_{rr}$ reaches to its peak at $t_2$ as shown in Fig. \ref{fig:reverse}.

\begin{figure}[t!]
\hspace{-0.5cm}
\includegraphics[scale=0.76]{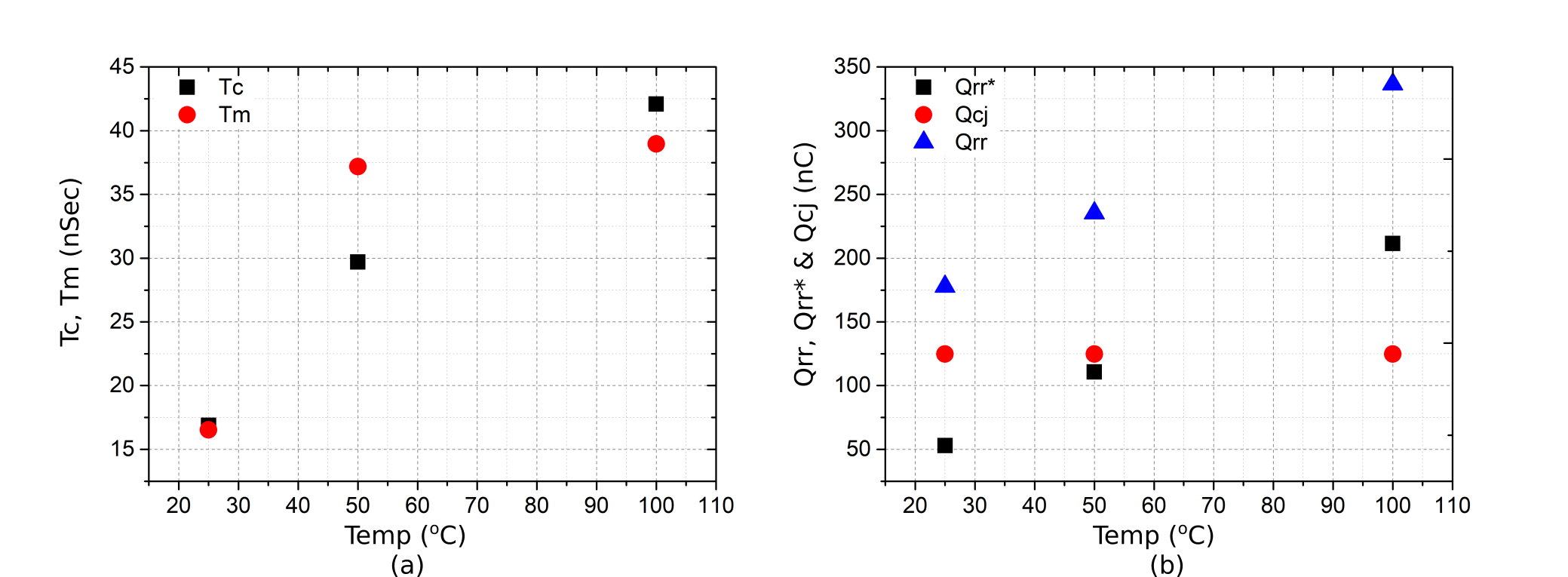}
\setlength{\abovecaptionskip}{-2mm}
\setlength{\belowcaptionskip}{0mm}
\caption{ (a)Variation of $T_m$ and $\tau_{c}$ with respect to temperature ($T_{j}$) and (b)Variation of $Q_{rr}$, $Q_{rr}^{*}$ and $Q_{cj}$ with respect to temperature ($T_{j}$)}
\label{fig:tctmqrr}
\end{figure}

\par
The body-diode's time constants $\tau_{c}$ and $T_{m}$ are found out as per the discussion in Section III-B and are shown in Fig. \ref{fig:tctmqrr}(a). The increment in $\tau_{c}$ and $T_{m}$ with respect to $T_{j}$ increases rapidly at lower temperature range but tends to reduce at higher temperature Fig. \ref{fig:tctmqrr}(a). The change in $Q_{rr}$, $Q_{rr}^{*}$ and $Q_{cj}$ with respect to temperature is shown in Fig. \ref{fig:tctmqrr}(b). The dynamics of $\tau_{c}$ and $T_{m}$ and $Q_{rr}^{*}$ from the experimental results can be expressed as per eq. \ref{eqn:irr1}\ and \ref{eqn:irr2}:

\begin{equation}
\label{eqn:irr1}
\begin{split}
\tau_{c}(nSec) &= \alpha_1 T_{j}^{\beta_1} + \gamma_1 \\
T_{m}(nSec) &= \alpha_2 T_{j}^{\beta_2} + \gamma_2 \\ 
\end{split}
\end{equation}

\vspace{-2mm}

\begin{equation}
\label{eqn:irr2}
\begin{split}
Q_{rr}(nC) &= m_1 T_{j} + c_1 \\
Q_{rr}^{*}(nC) &= m_2 T_{j} + c_2
\end{split}
\end{equation}

The fitting parameters for the above equations are given in Table \ref{tab:table3}.

\begin{table}[t!]
\centering
\caption{Capacitance Curve Fitting Parameters}
\begin{center}
\begin{tabular}{c|c|c|c}
\specialrule{.1em}{.05em}{.05em}
\specialrule{.1em}{.05em}{.05em}
\textbf{$Capacitance$} & \textbf{$C_{o,x}$} & \textbf{$a_{x}$} & \textbf{$b_{x}$}\\
\hline
 $C_{oss}$	&	1040 pF		&	3	&	1.25	\\
\hline
 $C_{rss}$	&	344.2 pF	&	0.19&	1.25	\\
\specialrule{.1em}{.05em}{.05em}
\end{tabular}
\label{tab:table1}
\end{center}
\end{table}


\begin{table}[t!]
\caption{Non Linear Capacitances}
\begin{center}
\begin{tabular}{c | c | c | c }
\specialrule{.1em}{.05em}{.05em}
\specialrule{.1em}{.05em}{.05em}
\multirow{3}{*}{\textbf{Conditions}}
  & \textbf{$v_{gs}<V_{th}$}			& \textbf{$v_{gs}>V_{th}$}	 		& \textbf{$v_{gs}>V_{th}$}				\\
  & \textbf{$V_{ds}\geq v_{gs}-V_{th}$}	& \textbf{$V_{ds}> v_{gs}-V_{th}$}	& \textbf{$V_{ds}\leq v_{gs}-V_{th}$}	\\
  & \textbf{$Cutoff$}					& \textbf{$Saturation$}				& \textbf{$Ohmic$}						\\
\hline
\multirow{1}{*}{$C_{gd}$} 
  & $C_{gd,l}=8pF$						& $C_{gd,av}=12pF$			 		& $C_{gd,h}=500pF$						\\
\hline
\multirow{1}{*}{$C_{ds}$} 
  & $C_{ds,l}=72pF$						& $C_{ds,av}=111pF$			 		& $C_{ds,h}=540pF$						\\
\hline
\end{tabular}
\label{tab:table2}
\end{center}
\end{table} 


\begin{table}[t!]
\centering
\caption{Curve Fitting Parameters From Reverse Recovery Test}
\begin{center}
\begin{tabular}{c|c|c|c|c|c}
\specialrule{.1em}{.05em}{.05em}
\specialrule{.1em}{.05em}{.05em}
\textbf{$Parameters$} & \textbf{$\alpha$} & \textbf{$\beta$} & \textbf{$\gamma$} & \textbf{$m$} & \textbf{$c$}\\
\hline
 $\tau_{c}(nSec)$	&	-447.6		&	-0.04924	&	398.9		&	-		&	-		\\
\hline
 $T_{m}(nSec)$		&	-2.074e6	&	-3.55		&	39.12		&	-		&	-		\\
\hline
 $Q_{rr}(nC$		&	-			&	-			&	-			&	2.101	&	127.3	\\
\hline
 $Q_{rr}^{*}(nC)$	&	-			&	-			&	-			&	2.101	&	2.54	\\
\specialrule{.1em}{.05em}{.05em}
\end{tabular}
\label{tab:table3}
\end{center}
\end{table}

\begin{figure}[h!]
\centering{\includegraphics[scale=1.35]{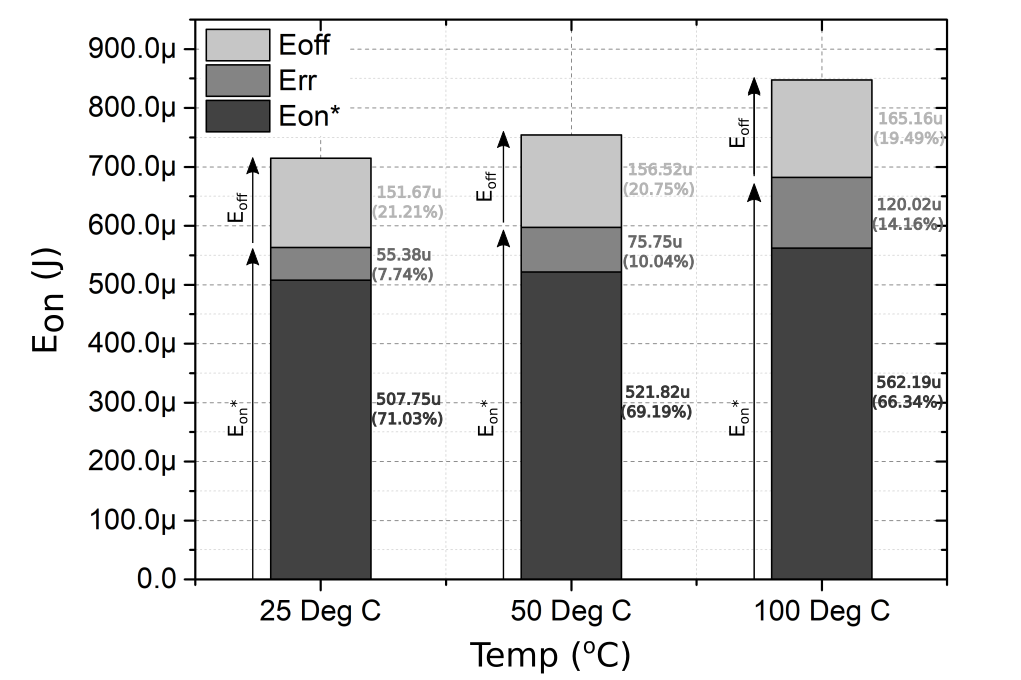}}
 \setlength{\abovecaptionskip}{-2mm}
\caption{ Analytical energy loss segments under $T_{j}$  = $25^{o}C$, $50^{o}C$ and $100^{o}C$ for $V_{dc}=600V$ and $I_{load} = 25A$}
\label{fig:etotvstj2}
\end{figure}

\subsection{Energy Loss Verification}
\par
The loss estimation model presented in Section III is used for computing the turn-on energy losses at different operating temperatures ($T_{j}$ $=$ $25^{o}C$, $50^{o}C$ and $100^{o}C$). The experimental energy losses are calculated by multiplying the $V_{ds2}$ and $I_{d2}$ during switching (Fig. \ref{fig:all}(e) and (h) for turn-on and Fig. \ref{fig:all}(d) and (g) for turn-off), and calculating the area under the curve. The analytical and experimental results of turn-on loss and total switching loss are compared and plotted in Fig. \ref{fig:all}(f) and (i) respectively. Over a wide operating range the maximum and minimum error in the estimation of $E_{on}$ are $+4.0\%$ and $-10.86\%$ respectively. From Fig. \ref{fig:all}(f) it can be observed that, due to the inclusion of temperature dependent SiC MOSFETs body-diode time constants, the proposed model's accuracy for $E_{on}$ calculation significantly increases at higher $T_{j}$ compared to the loss model in \cite{christen_analytical_2019}. The loss model in \cite{christen_analytical_2019} is used to plot the turn-on loss in Fig. \ref{fig:all}(f) under different loading conditions, using device data-sheet parameters which are available only at $T_{j}=25^oC$. It can be observed that due to non-inclusion of temperature dependent variations, the loss model in \cite{christen_analytical_2019} shows significant error at high temperature. This error has been shown to reduce with the proposed work in this paper.

\par
For calculating the total energy loss $E_{tot}$, estimation of turn-off energy loss $E_{off}$ is directly included, as per the method given in \cite{christen_analytical_2019} as $E_{off}$ does not change significantly with respect to $T_{j}$. Fig. \ref{fig:etotvstj2} shows the loss distribution in $E_{tot}$ and the energy loss due to reverse recovery of $Q_1$ on $Q_2$ is subtracted from eq. \ref{eqn:int1.8} and shown separately as $E_{rr}$. It can be seen that $E_{rr}$ increases by $116.7\%$, when $T_{j}$ is increased from $25^oC$ to $100^oC$. The absolute maximum and minimum error range in the estimation of $E_{tot}$ are $22.8\%$ and $4.0\%$, respectively. The error in $E_{tot}$ increases at low value of $I_{load}$ due to the estimation error in $E_{off}$ at $0$ $\leq$ $I_{load}$ $\leq$ $2I_{C}$. At this current level, $I_{ch2}$ of $Q_2$ becomes zero, due to which, $dV_{ds2}/dt$ and $I_{d2}/dt$ decreases significantly. In the experimental calculation, an accuracy band of $\pm5\%$ is created for the measurement error introduced by the voltage and current probes.

\section{Conclusion}
In this paper, a combination of an analytical and behavioural model for estimating the temperature-dependent turn-on switching losses of a SiC MOSFET in a half-bridge configuration is presented. The impact of displacement current due to the SiC MOSFETs parasitic capacitance on estimating the reverse-recovery charge of the body-diode has been analysed. The impact of temperature on the carrier lifetime of the body diode has been discussed. A parameter extraction routine has been followed for modelling the SiC MOSFET and its body diode characteristics. Further, the temperature-dependent charge carrier lifetime of SiC MOSFETs' body-diode has been experimentally calculated and used in the turn-on switching loss model. The proposed model can give the segmented analytical expression for turn-on energy losses, including the device non-idealities. For the validation of the model double-pulse test has been conducted at different $T_{j}$ ($25^oC, 50^oC \ \textrm{and} \ 100^oC$). The experimental and analytical results are compared over a wide operating range, and the absolute error in the estimation of turn-on switching energy loss is found to be  $4\%$ to $10.86\%$, respectively.

\bibliography{conference_041818}

\end{document}